\documentclass[prd,twocolumn,floats,floatfix,showpacs,nofootinbib,10pt]{revtex4}
\usepackage{graphicx}
\usepackage{dcolumn}
\usepackage{bm}
\usepackage{graphics}
\usepackage{slashed}
\usepackage{amssymb}
\usepackage{natbib}
\usepackage{amsmath}
\usepackage{url}
\usepackage{color}
\newcommand{\bea}{\begin{eqnarray}}
\newcommand{\ena}{\end{eqnarray}}
\newcommand{\bean}{\begin{eqnarray*}}
\newcommand{\enan}{\end{eqnarray*}}

\begin{document}

\title{Electromagnetic Geometry}
\author{M. Novello\footnote{M. Novello is Cesare Lattes ICRANet Professor: novello@cbpf.br}, F. T. Falciano\footnote{ftovar@cbpf.br},
E. Goulart\footnote{egoulart@cbpf.br}}
\affiliation{Instituto de Cosmologia Relatividade Astrofisica ICRA -
CBPF\\ Rua Dr. Xavier Sigaud, 150, CEP 22290-180, Rio de Janeiro,
Brazil}

\date{\today}

\begin{abstract}
We show that Maxwell's electromagnetism can be mapped into the Born-Infeld theory in a curved space-time, which depends only on the electromagnetic field in a specific way. This map is valid for any value of the two lorentz invariants $F$ and $G$ confirming that we have included all possible solutions of Maxwell's equations. Our result seems to show that specifying the dynamics and the space-time structure of a given theory can be viewed merely as a choice of representation to describe the physical system.
\end{abstract}

\maketitle

\section{Introduction}

The mathematical description of relativistic fields is grounded on two fundamental cornerstones: i) first, in the space-time metrical structure, which should come from Einstein's equations. and ii) in the specific field's lagrangian, which uniquely characterizes the field's evolution in the above-mentioned geometry. If gravity is negligible, general relativity predicts that the field propagates in a Minkowskian background with metric given by $\eta_{\mu\nu}$, which yields a simplified scenario from a computational point of view. Indeed, the behavior of fields in this simple flat spacetime has been the starting point of any field theoretical construction since the early days of relativity.

In this paper we discuss some arbitrariness concerning the combined notions of field dynamics and metrical structure. Typically, the equation of motion of any field (say, the electromagnetic field) is an intricate amalgam between the underlying metric structure and field derivatives. Specifying the metric tensor guarantees a well defined dynamics if the action functional is given. Nevertheless, it seems that there exist yet some degree of ambiguity in the characterization of the underlying background geometry. The reason is the following: given a background and an equation of motion, it is possible to simultaneously deform both (typically in a nonlinear fashion) in order to preserve the solutions of the former intact. In other words, specifying the dynamics and the space-time structure of a given theory may be viewed alternatively as a choice of representation (see \cite{N1} and \cite{N2}  for a complementary discussion). We are going to explore such different representations in the context of electrodynamics.

Indeed, we will show that Maxwell's linear electrodynamics in flat spacetime may be deformed into a Born-Infeld theory in a curved spacetime in terms of a prescribed map. This means that \textit{any} solution of the former will also be a solution of the latter. This highly nontrivial task becomes possible only if the curved spacetime depends explicitly on the electromagnetic field. Due to the algebraic structure of the electromagnetic two-form $F_{\mu\nu}$ and its dual, there exist a kind of closure relation that allows the existence of this map, hence, generating a ``dynamical bridge" between these two paradigmatic theories. We furnish the complete recipe of such map and discuss some of its fundamental properties.

At first, it may seems implausible that one could represent the simple linear Maxwell electrodynamics in terms of a much more involved non-linear Born-Infeld theory by using a modification of the metric structure. Nevertheless, we shall prove here that this equivalence is thorough and it is nothing but a matter of representation.

\section{Mathematical Setup}

We shall consider the source-free Maxwell's equations that can be written in its covariant form as
\begin{equation}
F_{ \phantom a \phantom a ; \nu}^{\mu \nu}=0 \quad\quad\stackrel{\ast}{F^{\mu\nu}}_{;\nu}=0 ,\label{max1}\\
\end{equation}
where the dual is given by $\stackrel{\ast}{F^{\mu\nu}}=\frac{1}{2} \eta^{\mu\nu}_{\phantom a \phantom a\alpha \beta}F^{\alpha \beta}$. With respect to a normalized congruence of observers $v^{\mu}$ we have
\begin{equation}
F^{\mu \nu}\equiv E^{\left[\mu\right.}v^{\left.\nu\right]}-\eta^{\mu \nu}_{\phantom a \phantom a \alpha \beta}H^{\alpha}v^{\beta}
\end{equation}
where the electric and magnetic fields are given by the projections $E^{\mu}\equiv F^{\mu}_{\phantom a \nu}v^{\nu}$ and $H^{\mu}\equiv \stackrel{\ast}{F^\mu_{\phantom a \beta}}v^{\beta}$. There are only two lorentz invariant quantities that can be constructed with the electromagnetic two-form, namely,
\begin{eqnarray*}
F&\equiv&F^{\mu \nu}F_{\mu \nu}=2\left(H^2-E^2\right)\\
G&\equiv&\stackrel{\ast}{F^{\mu \nu}}F_{\mu \nu}=-4\vec{E}.\vec{H}\\
\end{eqnarray*}
Using the above definitions, a direct calculation shows the following algebraic relations
\begin{eqnarray}
&&\stackrel{\ast}{F^{\mu \alpha}}\stackrel{\ast}{F_{\alpha \nu}}-F^{\mu \alpha}F_{\alpha \nu}=\frac{F}{2}\delta^\mu{}_{\nu}\label{algrel1}\\
&&\stackrel{\ast}{F^{\mu \alpha}}F_{\alpha \nu}=-\frac{G}{4}\delta^\mu{}_{\nu}\label{algrel2}\\
&&F^\mu{}_\alpha F^\alpha{}_\beta F^\beta{}_\nu=-\frac{G}{4}\stackrel{\ast}{F^\mu}{}_{\nu}-\frac{F}{2}F^\mu{}_{\nu}\label{algrel3}\\
&&F^\mu{}_\alpha F^\alpha{}_\beta F^\beta{}_\lambda F^\lambda{}_\nu=\frac{G^2}{16}\delta^\mu{}_{\nu}-\frac{F}{2}F^\mu{}_{\alpha}F^\alpha{}_{\nu} \quad \;\label{algrel4}
\end{eqnarray}
Note that, due to these identities, one can construct rank-2 objects only up to second power of $F^{\mu \nu}$, i.e. any power of the electromagnetic tensor and its dual is a combination of the identity $\delta^\mu{}_{\nu}\, , \, F^\mu{}_{\nu}$,  $\stackrel{\ast}{F^\mu}{}_{\nu}$ and $\phi^{\mu}{}_{\nu}\equiv F^\mu{}_{\alpha}F^\alpha{}_{\nu}$.

\subsection{Born-Infeld theory}

Originally, the Born-Infeld (BI) theory was an attempt to modify Maxwell's electromagnetism that could circumvent the self-energy divergence of the classical charged point-like particle \cite{b1}-\cite{Di}. Hence, one of its main motivations was to establish a consistent classical theory for the electron. Notwithstanding, the Born-Infeld theory has other interesting features that make it a distinguished theory among non-linear electromagnetic theories  \cite{Pleb}-\cite{Boi}. In the BI theory 1) excitations propagate without forming shocks that are common to generic nonlinear models; 2) there is a single characteristic surface equation, i.e. the birefringence phenomenon is absent; 3) it fulfills the positive energy density condition and 4) it also satisfies the duality invariance. Nowadays, there has been a renewed interest in BI theories and its non-abelian generalizations in connection with string theory or gauge fields on D-branes\cite{C}-\cite{Sk}, K-essence-like models in cosmology \cite{Gn}-\cite{Bab} and others. Thus, the BI is a good prototype of a nonlinear theory in the sense that it has general desirable properties from a physical point of view.

The BI lagrangian is defined\footnote{The constant term in the lagrangian has the purpose to make the energy-momentum tensor of the point-charge field goes to zero in the spatial infinity. In a cosmological scenario it can be interpreted as a kind of cosmological constant. However, here it plays no role whatsoever and can be omitted at will.} as
\begin{eqnarray*}\label{BI}
&&L=\beta^2\left(1-\sqrt{U}\right) \quad \mbox{with}\quad \beta=\mbox{Const}\\
&&U\equiv 1+\frac{F}{2 \beta^2}-\frac{G^2}{16 \beta^4}
\end{eqnarray*}

Variation with respect to the electromagnetic potential yields the following equation of motion
\begin{equation}
\left[\sqrt{-\gamma}\left(L_F F^{\mu \nu}+L_G \stackrel{\ast}{F^{\mu \nu}}\right)\right]_{,\nu}=0
\end{equation}
where $\gamma_{\mu \nu}$ is the Minkowski flat metric but in arbitrary coordinate system and $L_X$ is defined as the derivative of the lagrangian with respect to $X$. Note that in the limit $\beta\rightarrow\infty$ Maxwell's linear theory is recovered.

\section{Dynamical bridge: Maxwell into Born-Infeld}

Maxwell's electromagnetism in a flat Minkowski space-time is linear and hence, possibly the simplest classical field theory one can envisage. Notwithstanding, we will show that one can map this linear dynamics into a nonlinear structure defined by the Born-Infeld theory. The price to pay is to leave the Minkowski background $\gamma_{\mu\nu}$ and go to a specific  curved space-time $\hat{q}_{\mu\nu}$ that is constructed solely in terms of the background metric and the electromagnetic field. We call the map that implements such a modification of representation a ``dynamical bridge".

To describe a physical theory one has to specify not only the lagrangian that contains the dynamics of the fields but also the space-time structure where the theory is defined. To avoid notational cumbersomeness, we shall define every object in the curved space-time defined by the metric $\hat{q}_{\mu \nu}$ with an upper hat. Thus, in Maxwell's theory, every tensor is raised and lowered by the Minkowski metric $\gamma_{\mu \nu}$ while in the curved space-time representation, where we will define the Born-Infeld theory, tensors shall be raised and lowered using the $\hat{q}_{\mu \nu}$ metric. We have the following definitions\\\\
Maxwell Theory: $\left\{\gamma_{\mu \nu} \; ,\;  L=-\frac14 F\right\} $
\begin{eqnarray*}
&&F^{\mu \nu}=F_{\alpha \beta}\gamma^{\mu \alpha}\gamma^{\nu \beta}\\
&&\\
&&F=F_{\mu \nu}F_{\alpha \beta}\gamma^{\mu \alpha}\gamma^{\nu \beta}\qquad \qquad \qquad \qquad \\
\end{eqnarray*}
Born-Infeld Theory: $ \left\{ \hat{q}_{\mu \nu} \; , \; \hat{L}=\beta^2\left(1-\sqrt{\hat{U}}\right)\right\}$
\begin{eqnarray*}
&&\hat{U}=1+\frac{\hat{F}}{2\beta^2}-\frac{\hat{G}}{16\beta^4}\qquad \\
&&\\
&&\hat{F}^{\mu \nu}=F_{\alpha \beta}\hat{q}^{\mu \alpha}\hat{q}^{\nu \beta}\\
&&\\
&&\hat{F}=F_{\mu \nu}F_{\alpha \beta}\hat{q}^{\mu \alpha}\hat{q}^{\nu \beta}\\
&&\\
&&\hat{G}=\frac{\sqrt{-\gamma}}{\sqrt{-\hat{q}}}\eta^{\mu \nu \alpha \beta} F_{\mu \nu}F_{\alpha \beta}=\hat{F}{}^{\stackrel{\ast}{\mu\nu}}F_{\mu \nu}
\end{eqnarray*}

Let us define the metric $\hat{q}^{\mu \nu}$ that depends on the electromagnetic field. Due to the algebraic relations (\ref{algrel1})-(\ref{algrel4}), there is a unique way to introduce what we call the electromagnetic metric (EM metric), namely,
\begin{equation}\label{qmet}
\hat{q}^{\mu \nu}=a\gamma^{\mu \nu}+b\phi^{\mu \nu}
\end{equation}
where $a$ and $b$ are two arbitrary real functions of the two lorentz invariants $F$ and $G$ and again $\phi^{\mu \nu}\equiv F^{\mu \alpha}F_{\alpha}{}^{\nu}$. The term electromagnetic metric is justified by the fact that $\hat{q}_{\mu \nu}$ depends only on the electromagnetic field.

Note the very convenient property that the inverse of the above structure, which in general is given by an infinite series\footnote{Any metric $g^{\mu \nu}$ can be decomposed as $g^{\mu \nu}=\eta^{\mu \nu}+h^{\mu \nu}$ with $\eta^{\mu \nu}$ the Minkowski metric and $h^{\mu \nu}$ a non-negligible terms. By the definition of the inverse, i.e. $g^{\mu \alpha}g_{\alpha \nu}=\delta^\mu{}_\nu$, one can show that the inverse $g_{\mu \nu}$ is given as an infinite series as $g_{\mu \nu}=\eta_{\mu \nu}-h_{\mu \nu}+h_{\mu}{}^\alpha h_{\alpha \nu}+\ldots$.}, has only two terms. This is a consequence of the algebraic properties (\ref{algrel1})-(\ref{algrel4}). Indeed, given that the inverse satisfies $\hat{q}^{\mu \alpha}\hat{q}_{\alpha \nu}=\delta^\mu{}_\nu$, we have
\begin{equation}\label{qmet2}
\hat{q}_{\mu \nu}=A\gamma_{\mu \nu}+B\phi_{\mu \nu}
\end{equation}
with
\[
A=\frac{2-\varepsilon F}{2aQ}\quad,\qquad B=-\frac{\varepsilon}{aQ}\qquad ,
\]
and the following definitions
\[
\varepsilon \equiv \frac{b}{a} \quad , \quad Q\equiv 1-\frac{\varepsilon F}{2}-\frac{\varepsilon^2G^2}{16}\qquad .
\]

Through their definitions, one can relate the Born-Infeld fields in terms of the Maxwell's ones. For instance, the new invariants read
\begin{equation*}
\hat{F}=a^2\left[F-\varepsilon\left(F^2+\frac{G^2}{2}\right)+\frac{\varepsilon^2 F}{4}\left(F^2+\frac34 G^2\right)\right]\quad ,
\end{equation*}
\begin{equation*}
\hat{G}=a^{2}\left(1-\frac{\varepsilon F}{2}-\frac{\varepsilon^{2}G^{2}}{16}\right)G\quad .
\end{equation*}

At this stage, the mapping of the two dynamical systems sum up to make the identification term by term in both equations. In other words, choosing appropriately the unknown functions of the EM metric (\ref{qmet}) \textit{Maxwell's equation and Born-Infeld in this curved space-time will reproduce the same dynamics}. We start our task by comparing the first equation of (\ref{max1}) in the background $\gamma_{\mu\nu}$ with (\ref{BI}) in the background $\hat{q}_{\mu\nu}$.

Equating terms proportional to $F^{\mu \nu}$ and to $\stackrel{\ast}{F}{}^{\mu \nu}$ we find the following conditions to be fulfilled in order to obtain the equality of the dynamics
\begin{eqnarray}
&&\frac{a^2 Q^2}{2 \beta^2}= -\varepsilon + \frac{\varepsilon^2 F}{4} \quad \label{1}\\
&&1-\varepsilon F+\frac{\varepsilon^2 F^2}{4} +\frac{\varepsilon^2 G^2}{16}=Q\sqrt{\hat{U}} \quad  \qquad \label{2}
\end{eqnarray}

In the above equations we are considering $a$ and $\varepsilon$ as the unknown functions and $Q$ and $\hat{U}$ as functions of the formers. Having two equations and two unknown, one can solve the system to find the two functions, hence, the metric (\ref{qmet}). However, using equation (\ref{1}), it follows immediately that equation (\ref{2}) trivialize to $1=1$, i.e. if (\ref{1}) is satisfied then (\ref{2}) is also immediately satisfied. Consequently, if $a$ and $\varepsilon$ satisfy the constraint (\ref{1}), Maxwell's electromagnetism in Minkowski spacetime and Born-Infeld in the $\hat{q}_{\mu \nu}$ spacetime are completely equivalent. Note that the Bianchi identity in the electromagnetic metric $\hat{q}_{\mu\nu}$
\begin{equation}
(\sqrt{-\hat{q}}\hat{F}{}^{\stackrel{\ast}{\mu\nu}}),_{\nu}=0
\end{equation}
is identically satisfied. We would like to emphasize that although the two metrical structures are different, we are dealing with two representations of the same and unique dynamical process.

Beside the freedom in one of the functions that define the metric $\hat{q}_{\mu \nu}$, which seems to be related to the conformal invariance of Maxwell's equations, there are still some conditions that constrict these functions. A straightforward calculation using the Cayley Hamilton theorem shows that the determinant of the metric (\ref{qmet2}) gives
\[
\sqrt{-\hat{q}}=a^{-2}Q^{-1}\sqrt{-\gamma}\quad.
\]

Therefore, $Q$ that is a function of $\varepsilon$ has to be positive definite, i.e. $Q>0$. In addition, we want the map to be for any solution of Maxwell's equations, hence for any value of the two invariants $F$ and $G$. Let us investigate some particular regimes.
\begin{itemize}
\item[i)] \underline{case $F=0$}

In this case, equation (\ref{1}) demands that $\varepsilon<0$. Furthermore, by its own definition $\varepsilon =- \frac{4}{|G|}\sqrt{1-Q}$, which restrict $Q$ to the domain $0<Q\leq 1$.\\

\item[ii)] \underline{case $G=0$}\\

The relation between $\varepsilon$ and $Q$ now gives $\varepsilon=\frac{2}{F}\left(1-Q\right)$. In addition, from eq.(\ref{1}) we can solve for the function $a$ as
\[
a=\frac{\beta}{Q}\sqrt{\frac{2}{F}\left(Q^2-1\right)} \quad .
\]
Since $a$ has to be a real function, we have
\begin{displaymath}
G=0 \rightarrow \left\{\begin{array}{l}
F>0 \quad \rightarrow \; Q>1\\
F<0 \quad \rightarrow \; 0<Q<1
\end{array}\right.
\end{displaymath}

\item[iii)] \underline{case $F=G=0$}\\

The vanishing of both invariants simultaneously, which is the case for plane waves, trivialize all conditions. In fact, we simply have $Q=1$ and $a^2=-2\beta^2 \varepsilon$. Thus, it suffice to have $\varepsilon<0$.
\end{itemize}

\begin{figure}[h!]
\includegraphics[width=7.5cm,height=5cm]{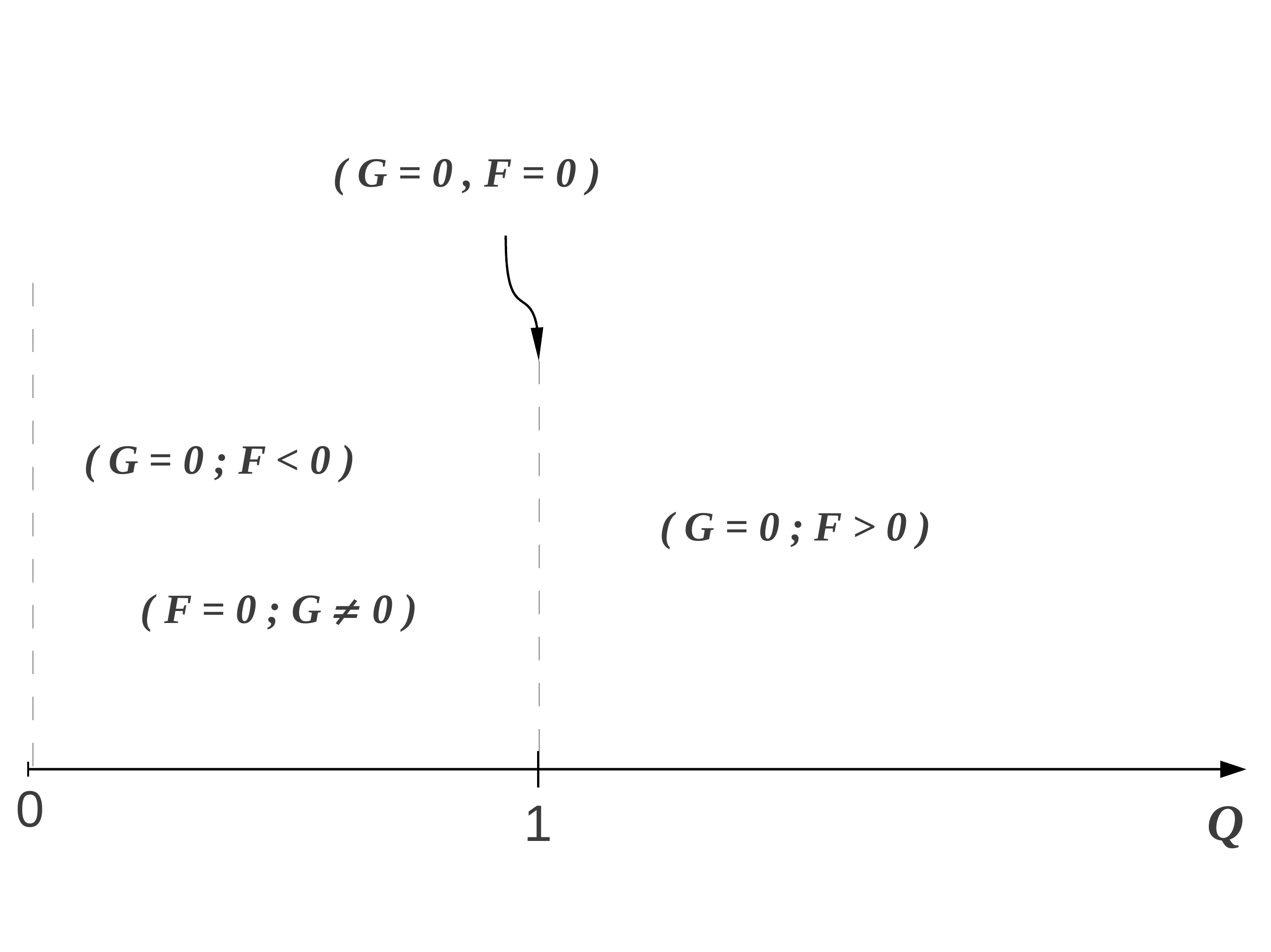}
\caption{\small{Range of dependence of the function $Q$ with respect to the two electromagnetic invariants $F$ and $G$.}}\label{pic1}
\end{figure}

The above conditions can be visually summarized in figure \ref{pic1}. The behavior of $Q$ going from less than 1 to bigger than 1 as $F$ changes sign is a typical behavior of exponential functions. The dependence in $G$ has broader options but being positive definite and always less than 1 again is typical of a hyperbolic secant or a function like $\left(1+G^2\right)^{-1}$. For $G\neq0$ we can solve for $\varepsilon$ and find
\[
\varepsilon=-\frac{4}{G^2}\left\{1+\sqrt{\Delta}\right\} \quad \mbox{with}\quad \Delta \equiv F^2+G^2(1-Q)\quad .
\]

A reasonable functional dependence for $Q$ that satisfies all the above conditions is
\begin{equation}\label{Q}
Q=\frac{e^{F/4 \beta^2}}{1+G^2 . \, e^{F/4 \beta^2}}
\end{equation}
We need to include the $\beta^2$ in the argument of the exponential to cancel the dimensionality of $F$. The denominator of the above expression was
defined so that if $G=0$ it goes to a pure exponential while if $G\neq 0$ then we are always in the range $\Delta \geq 0$. With ansatz (\ref{Q}) we guarantee that $\varepsilon$ and $a$ are always well defined for any value of the two invariants $F$ and $G$. As a result, we have shown that there exist a complete equivalence between the two representations, i.e. the Born-Infeld theory represented in the spacetime with the electromagnetic metric $\hat{q}_{\mu\nu}$ has the same physical content as Maxwell's theory in a Minkowski background.

\subsection{Electromagnetic metrics}

Having established the connection between Maxwell's electromagnetism and Born-Infeld theory, we can now analyze some examples of metrics generated by this map. Maybe the most interesting class of examples are the $G=0$, which includes the case of pure electrostatic or magnetostatic fields. Let us consider $G=0$, i.e.
\begin{eqnarray*}
Q=e^{F/4 \beta^2}\quad &\mbox{and}&\qquad a=\frac{\beta}{Q}\sqrt{\frac{2}{F}\left(Q^2-1\right)} \quad .
\end{eqnarray*}

There are three limiting cases: weak fields $|F|\ll 1$, strong electric field $F\ll -1$ and strong magnetic field $F\gg 1$.

\begin{itemize}
\item[I)] $|F|\ll 1$\\
For weak fields, expanding in powers of $F$ we find
\begin{eqnarray*}
&&Q \approx 1+\frac{F}{4\beta^2}+\frac{F^2}{32\beta^4}+\mathcal{O}\left(\frac{F^3}{\beta^6}\right) \quad ,\\
&&a \approx 1-\frac{F}{8\beta^2}+\frac{5F^2}{384\beta^4}+\mathcal{O}\left(\frac{F^3}{\beta^6}\right)\quad,
\end{eqnarray*}
hence, we have in first order the metric
\begin{equation}\label{qi}
\hat{q}_{\mu \nu}=\gamma_{\mu \nu}+\frac{1}{2\beta^2}F_{\mu}{}^{\alpha}F_{\alpha \nu}
\end{equation}

\item[II)] $F\ll -1$\\
Considering $|F|\gg 1$ and $F<0$, we have
\begin{eqnarray*}
Q=e^{F/4 \beta^2}\quad &\mbox{and}& a \approx \beta\sqrt{\frac{2}{|F|}}e^{-F/4 \beta^2}\quad,
\end{eqnarray*}
hence, the metric assumes the form
\begin{eqnarray}
\hat{q}_{\mu \nu}&=&\sqrt{\frac{|F|}{2\beta^2}}e^{F/4 \beta^2}\gamma_{\mu \nu}+\frac{1}{\beta}\sqrt{\frac{2}{|F|}}F_{\mu}{}^{\alpha}F_{\alpha \nu}\nonumber\\
&\approx& \frac{1}{\beta}\sqrt{\frac{2}{|F|}}F_{\mu}{}^{\alpha}F_{\alpha \nu}\quad \label{qii}
\end{eqnarray}

\item[III)] $F\gg 1$\\
The case with $F\gg 1$ give us
\begin{eqnarray*}
Q=e^{F/4 \beta^2}\quad &\mbox{and}& a \approx \beta\sqrt{\frac{2}{F}}\quad,
\end{eqnarray*}
which yields a metric of the form
\begin{equation}\label{qiii}
\hat{q}_{\mu \nu}=\sqrt{\frac{F}{2\beta^2}}\gamma_{\mu \nu}+\frac{1}{\beta}\sqrt{\frac{2}{F}}F_{\mu}{}^{\alpha}F_{\alpha \nu}
\end{equation}
\end{itemize}

These are only simple illustrative examples. But it is worth noting that metric (\ref{qi}) for weak fields ($|F|\ll1$) has the same form as the newtonian limit in general relativity. We shall analyze each one of these examples in more detail and consider some possible applications in future works.

\section{Conclusion}

Let us stress that a given theory is only properly defined when one specifies together with the dynamical fields and their lagrangian the background environment, i.e. the space-time structure where the theory is defined.

In the present work, we have shown that Maxwell's electromagnetism can be mapped into a Born-Infeld theory in curved space-time. The electromagnetic metric of the Born-Infeld theory, namely eq.(\ref{qmet}), is unique in form and depends only on the electromagnetic field. We have proved the equivalence between these two theories for a source-free system. Next, we shall generalize our result to include a source term. Note that this is a non-trivial step inasmuch, in this case, the map should also consider the coupling between matter and the electromagnetic field.

It is remarkable that a linear theory as simple as Maxwell's electromagnetism can be recast in a non-linear theory in curved space-time. One could argue that being these two theories equivalent, we should keep the simplest and don't even bother with this correspondence. However, it must be noted that once established their equivalence, one could use this result in the inverse sense, i.e. to go from a non-linear to a linear dynamics.

Furthermore, the manifest dependence of the dynamics of a given theory on the space-time structure acquires a different meaning with our result. Not only the quest for the real nature of the space-time becomes ambiguous but it also defies the very definition of the correct dynamics that describes the physical phenomena. It seems that only a combination of the dynamics together with the spacetime structure shows to have physical meaning. We shall come to these matters in the near future.


\section{acknowledgements}
M. Novello would like to thank FINEP, CNPq and FAPERJ, E. Goulart FAPERJ for their financial support.



\begin{thebibliography}{50}
\bibitem{N1} M Novello and E Goulart, Class. Quantum Grav. 28 145022 (2011)

\bibitem{N2} E. Goulart, M. Novello, F. T. Falciano, J. D. Toniato, arXiv:1108.6067v1 [gr-qc] (2011), accepted for publication in Class. Quantum Grav.

\bibitem{b1} M. Born and L. Infeld, Nature 132, 1004 (1933).

\bibitem{b2} M. Born and L. Infeld, Proc. Roy. Soc. (London) 144, 425 (1934).

\bibitem{Sc0} E. Schrodinger, Proc. Roy. Soc. 150A (1935) 465.

\bibitem{Sc} E. Schrodinger, Proc. R. I. A, vol. XLVII, Sect A., 77 (1942)

\bibitem{Di}P. A. M. Dirac, Royal Society of London Proceedings Series A 268 (June, 1962) 57–67.


\bibitem{Pleb} J. Plebansky, in Lectures on Non-Linear Electrodynamics, (Ed. Nordita, Copenhagen, 1968).

\bibitem{Boi} G. Boillat, J. Math. Phys. 11, 941 (1970).

\bibitem{Boi2} G.Boillat, C. R. Acad. Sci. Paris 262 (1966) 1285.

\bibitem{Ker} R. Kerner, A.L. Barbosa∗ and D.V. Gal’tsov, arXiv:hep-th/0108026v2 (2001)

\bibitem{Bia} Z. Bialinicka-Birula and I. Bialinicki-Birula, Phys. Rev. D 2, 2341 (1970).

\bibitem{Bia2}  I. Bialinicki-Birula, Acta Physica Polonica B, vol. 30, no 10, 2875 (1999)

\bibitem{Y} Y. Brenier, Arch. Rational Mech. Anal. 172, 65 (2004).

\bibitem{C} C.Callan, J.Maldacena, Brane Dynamics From the Born-Infeld Action, Nucl. Phys. B513 (1998) 198, hep-th/9708147.

\bibitem{G1} G.W.Gibbons, D.Rasheed, Electric-magnetic duality rotations in non-linear electrodynamics, Nucl. Phys. B454 (1995) 185.

\bibitem{G2} G.W.Gibbons, Born-Infeld particles and Dirichlet p-branes, Nucl. Phys. B514 (1998) 603.

\bibitem{G3} G. W. Gibbons and C. A. R. Herdeiro,Phys. Rev. D 63, 064006 (2001).

\bibitem{Bau} D. Baumann et al., On d3-brane potentials in compactifications with fluxes and wrapped d-branes, JHEP 11 (2006) 031 [hep-th/0607050].

\bibitem{Sk} S. Kachru, R. Kallosh, A. Linde and S. P. Trivedi, De sitter vacua in string theory, Phys. Rev. D68 (2003) 046005 [hep-th/0301240].

\bibitem{Bro} A. R. Brown, S. Sarangi, B. Shlaer and A. Weltman, arXiv:0706.0485 [hep-th].

\bibitem{Gn} G. N. Felder, L. Kofman and A. Starobinsky, Caustics in tachyon matter and other born-infeld scalars, JHEP 09 (2002) 026 [hep-th/0208019].

\bibitem{Fro} A. V. Frolov, L. Kofman and A. A. Starobinsky, Prospects and problems of tachyon matter cosmology, Phys. Lett. B545 (2002) 8–16 [hep-th/0204187].

\bibitem{Lind} J. E. Lidsey and I. Huston, Gravitational wave constraints on dirac-born-infeld inflation, JCAP 0707 (2007) 002 [arXiv:0705.0240 [hep-th]].

\bibitem{Bab} E. Babichev, V. Mukhanov, A. Vikman, JHEP 0802 (2008) 101, 0708.0561 [hep-th].


\end{thebibliography}
\end{document}